\documentclass[aps,prb,reprint,showpacs,amssymb,amsmath,longbibliography]{revtex4-1}
\usepackage{graphicx}

\def\be{\begin{equation}}
\def\ee{\end{equation}}
\def\ber{\begin{eqnarray}}
\def\eer{\end{eqnarray}}
\def\bern{\begin{eqnarray*}}
\def\eern{\end{eqnarray*}}

\def\rv{\mathbf{r}}

\def\kv{\mathbf{k}}

\def\qv{\mathbf{q}}

\def\Rv{\mathbf{R}}

\def\0v{\mathbf{0}}
\def\1v{\mathbf{1}}
\def\2v{\mathbf{2}}
\def\3v{\mathbf{3}}

\def\ii{\mathrm{i}}

\begin{document}

\title{Exact ``exact exchange"  potential of  two- and one-dimensional electron gases
beyond the asymptotic limit}
\author{Vladimir~U.~Nazarov}
\affiliation{Research Center for Applied Sciences, Academia Sinica, Taipei 11529, Taiwan}
\email{nazarov@gate.sinica.edu.tw}

\begin{abstract}
The exchange-correlation potential experienced by an electron 
in the free space adjacent to a solid surface or to a low-dimensional system defines the fundamental image states
and is generally important in surface- and nano-science.
Here we determine  the potential
near the two- and one-dimensional electron gases (EG), doing this analytically  
at the level of the exact exchange of the density-functional theory (DFT). 
We find that, at $r_\perp\gg k_F^{-1}$, 
where  $r_\perp$ is the distance from the EG and $k_F$ is the Fermi radius, 
the potential obeys the already known asymptotic $-e^2/r_\perp$,
while at $r_\perp \lesssim k_F^{-1}$, but {\em still in vacuum}, 
qualitative and quantitative deviations of the exchange potential from the asymptotic law occur. 
The playground of the excitations to the low-lying image states
falls into the latter regime, causing significant departure from the Rydberg series.
In general, our analytical exchange potentials establish benchmarks for numerical approaches
in the low-dimensional science, where DFT is by far the most common tool. 
\end{abstract}


\maketitle

\section{Introduction}
The image potential (IP)  -- a potential experienced by a  test charge
outside a semi-infinite medium, a slab, or a system of a lower dimensionality -- is a
fundamental concept of classical electrostatics \cite{Landau-60}.
It is widely believed, although never proven \cite{Lang-70,Dobson-95,Constantin-11}, that 
the exchange-correlation (xc) part of the Kohn-Sham (KS)
potential of the density-functional theory (DFT) \cite{Kohn-65}
must asymptotically reproduce the classical IP at large distances from an extended system.
Much effort has been exerted over years to describe IP quantum-mechanically,
both in order 
to account for the experimentally important image states 
at solid surfaces and at quasi-low-dimensional systems, 
and to gain better understanding of the non-trivial interrelations between DFT and classical physics
\cite{Sham-85,Harbola-87,Eguiluz-92,Horowitz-06,Horowitz-10,Constantin-11,Qian-12,Engel-14,Engel-14-2}.   

Regardless of the ultimate answer to the question of whether or not the xc potential (which is not a physical quantity)
is equal in vacuum to the IP for a test charge (which {\em is} a physical quantity) 
\footnote{See Appendix \ref{CLASS} for the classical image potential of 2(1)DEG.}, 
the determination of the former is fundamentally important in quantum physics.
Indeed, it defines the KS band-structure of the system of interest, which step, followed
by a calculation of the system's response within the time-dependent DFT \cite{Zangwill-80,Runge-84,Gross-85,Gross-86e},
will produce excitations to image states (which are quite physical properties).
While for semi-infinite media the problem still remains  highly controversial 
\cite{Dobson-95,Constantin-11}, for {\em slabs} it has been firmly established \cite{Horowitz-06,Engel-14} 
that, on the level of the optimized effective potential-exact exchange (OEP-EXX) \cite{Sharp-53,Talman-76,Gorling-94}, the KS potential
has the asymptotic $-e^2/r_\perp$, valid at large distances $r_\perp$ from the slab.
However, apart from the thickness $a$, EG in the shape of a slab (quasi-2D EG) or 
a cylinder (quasi-1D EG), when considered quantum-mechanically, has a fundamental
intrinsic parameter  $k_F$ -- the Fermi radius -- and, therefore,
even at $a=0$, two different regimes,  at $r_\perp \gg k_F^{-1}$ and 
$r_\perp \lesssim k_F^{-1}$, can be anticipated.

At variance with a vast literature on 
the asymptotic behaviour of the xc potential, in this paper
we are concerned with the  potential in the whole space
outside a 2(1)DEG. 
We solve this problem {\em exactly} and {\em analytically} at the EXX level of DFT
and find that,
at $r_\perp \lesssim k_F^{-1}$, but still in vacuum, the potential is qualitatively and quantitatively 
different from its asymptotic form $-e^2/r_\perp$.
However, at larger distances $r_\perp \gg k_F^{-1}$, our potentials obey the correct asymptotic, which is known
to be mandotary for slabs in general \cite{Engel-14-2}. 
The non-asymptotic shape of the potential in the $r_\perp \lesssim k_F^{-1}$  region  strongly affects experimentally important 
low-lying image states, causing  significant deviations from the Rydberg series.

This paper is organized as follows: In Sec.~\ref{QUASI} we derive a closed-form 
EXX potentials for quasi-2(1)DEG with one filled subband. In Sec.~\ref{STRICT} we take the full confinement limit,
obtaining analytical EXX potentials for 2D and 1D electron gases. In Sec.~\ref{DISC} we visualize and discuss the results.
In Sec. \ref{LHF} we derive further insights from addressing the problem of quasi-2(1)DEG within the localized Hartree-Fock method.
Section \ref{CONCL} contains conclusions.
In Appendix \ref{CLASS} we discuss the classical image potential in two and one dimensions.
In Appendix \ref{FURTH}, finer details of the derivation of the main results are given.
Atomic units ($e^2=\hbar=m_e=1$) are used throughout.

\begin{figure} [h] 
\includegraphics[width=   \columnwidth, trim= 0 0 30 4, clip=true]{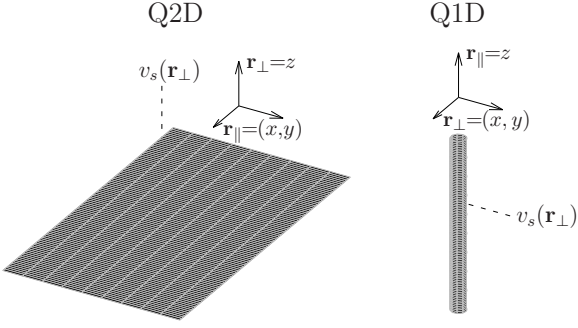}
\vspace{-0.5 cm}
\caption{\label{sys}
Schematics of quasi-2D (left) and quasi-1D (right) electron gases
and the notations adopted.
}
\end{figure}

\section{EXX potential of quasi-2(1)D electron gas with one subband filled}
\label{QUASI}

We start by considering  a quasi $d$-dimensional ($d=2,1$), generally speaking, spin-polarized EG. 
The positively charged background with the $d$-dimensional density $n$ is strictly
confined to the $xy$-plane and to the $z$-axis, for $d=2$ and $d=1$, respectively.
We are concerned with the KS problem \cite{Kohn-65} 
for spin-orbitals $\psi^\sigma_i(\rv)$ and eigenenergies $\epsilon^\sigma_i$, with the potential
\begin{equation}
v^\sigma_s(\rv)=v_{ext}(\rv)+v_H(\rv)+v^\sigma_{xc}(\rv),
\label{vs}
\end{equation}
generally speaking, different for different spin orientations $\sigma$,
where $v_{ext}(\rv)$, 
$v_H(\rv)=\int n(\rv')/|\rv-\rv'| d\rv'$,
and 
\begin{equation}
v^\sigma_{xc}(\rv)=\frac{\delta E_{xc}}{\delta n^\sigma(\rv)}
\label{vxc}
\end{equation}
are the external, Hartree, and xc potentials, respectively,
$n^\sigma(\rv)$ are spin-densities, $n(\rv)=n^\uparrow(\rv)+n^\downarrow(\rv)$ is the particle-density,
and $E_{xc}$ is the xc energy. For the latter we  use the EXX part,
which can be written as \cite{Slater-51}
\begin{equation}
E_x= - \frac{1}{2} \int \frac{|\rho(\rv,\rv')|^2}{|\rv-\rv'|} d\rv d\rv',
\label{Exg}
\end{equation}
where
\begin{equation}
\rho(\rv,\rv') = \sum\limits_{i\in occ} \phi_i(\rv) \phi_i^*(\rv') 
\label{ro}
\end{equation}
is the density-matrix, the summation in Eq.~(\ref{ro}) running over the occupied states only.

Let $\rv_\|$ and 
$\rv_\perp$ be the coordinate vectors parallel and perpendicular, respectively, to the extent of the EG.
In other words, for $d=2$, $\rv_\|$ is the radius-vector in the $xy$ plane and $\rv_\perp$ is a vector in the $z$-direction,
while for $d=1$  this is vice versa, as schematized in Fig.~\ref{sys}. 
Keeping in mind the subsequent zero-thickness limit,
we assume  one, at most, subband with the wave-function
$\mu^\sigma_0(\rv_\perp)$ to be occupied for each spin direction.
Therefore, all spin-orbitals with the same spin orientation have the same $\rv_\perp$-dependence
\begin{equation}
\phi_{\kv_\|}^\sigma(\rv)=\frac{e^{\ii \kv_\| \cdot \rv_\|}}{\Omega^{1/2}} \mu^\sigma_0(\rv_\perp),
\label{DMF}
\end{equation}
where $\Omega$ is the normalization area or length, in the 2D and 1D cases, respectively.
Then, by Eq.~(\ref{ro}), for the density-matrix we can write 
\begin{equation}
\rho(\rv,\rv') = \sum\limits_\sigma
\mu^\sigma_0(\rv_\perp) {\mu^\sigma_0}^*(\rv_\perp') \rho^\sigma(\rv_\|-\rv'_\|),
\label{rrp}
\end{equation}
where 
\begin{equation}
\rho^\sigma(\rv_\|) =
\frac{  1}{\Omega}  \sum\limits_{|\kv_\||\le k^{\sigma}_F} \! \! e^{\ii \kv_\| \cdot \rv_\|},
\label{rofrp}
\end{equation}
$k^\sigma_F$ being the Fermi radii for the corresponding spin orientations.
As will be seen below, the factorization (\ref{rrp}) 
is the key property of 2(1)DEG with only one subband occupation,
which makes possible the explicit solution to the EXX problem for these systems.
Then, the exchange energy of Eq.~(\ref{Exg})  can be written as
\begin{equation}
\begin{split}
E_x=-\frac{1}{2} \sum\limits_\sigma \int \frac{|\mu^\sigma_0(\rv_\perp)|^2 |\mu^\sigma_0(\rv_\perp')|^2 
|\rho^\sigma(\rv_\|  -  \rv'_\|)|^2}{|\rv-\rv'|} d\rv d\rv' \\
=-\frac{1}{2} \sum\limits_\sigma \int \frac{n^\sigma(\rv) n^\sigma(\rv') 
|\rho^\sigma(\rv_\|  -  \rv'_\|)|^2}{(n^\sigma)^2 |\rv-\rv'|} d\rv d\rv',
\end{split}
\label{Ex2}
\end{equation}
where the spin-density is
\begin{equation}
n^\sigma(\rv) = \rho^\sigma(\rv,\rv)=
|\mu^\sigma_0(\rv_\perp)|^2 n^\sigma,
\label{nnp}
\end{equation}
and $n^\sigma=\int n^\sigma(\rv) d\rv_\perp$ is the $d$-dimensional uniform  spin-density.
By virtue of Eqs.~(\ref{vxc}),  (\ref{Ex2}), and (\ref{nnp}), the  exchange potential
evaluates explicitly to
\footnote{In Eq.~(\ref{m1fin1}), the 3D spin-density $n^\sigma(\rv)$ is varied
at a fixed value of the $d$-dimensional spin-density $n^\sigma$ (and, therefore, $k^\sigma_F$), which amounts
to the conservation of the total number of particles. 
In Appendix \ref{FURTH} further particulars of the derivation are given.
For an alternative proof of Eq.~(\ref{m1fin1}) in terms of the KS eigenfunctions and eigenenergies, see Appendix \ref{FURTH2}.}.
\begin{equation}
\begin{split}
v^\sigma_x(\rv) &  \! = \! \frac{\delta E_x}{\delta n^\sigma(\rv)} \! = \!
- \frac{1}{(n^\sigma)^2} \! \int \! \frac{n^\sigma(\rv') 
|\rho^\sigma(\rv_\|  \! - \!  \rv'_\|)|^2}{ |\rv-\rv'|} d\rv'
\\ &=
- \frac{1}{n^\sigma} \int \frac{|\mu^\sigma_0(\rv'_\perp)|^2 |\rho^\sigma(\rv_\|-\rv'_\|)|^2}{|\rv-\rv'|} 
d \rv'. 
\end{split}
\label{m1fin1}
\end{equation}
In Eq.~(\ref{m1fin1}) we easily recognize the Slater's exchange potential \cite{Slater-51}.
This leads us to the important conclusion
that, for  Q2(1)EG with {\em only one subband filled} for each spin component,
EXX and the Slater's potentials coincide exactly and, consequently, in this case, the  EXX potential
can be expressed in terms of the occupied states only.
The latter becomes wrong when more subbands are filled
\footnote{In particular, this does not hold for a metal surface, where the Slater and OEP potentials differ \cite{Harbola-87}}.

\begin{figure} [h] 
\includegraphics[width=  \columnwidth, trim= 25 0 30 0, clip=true]{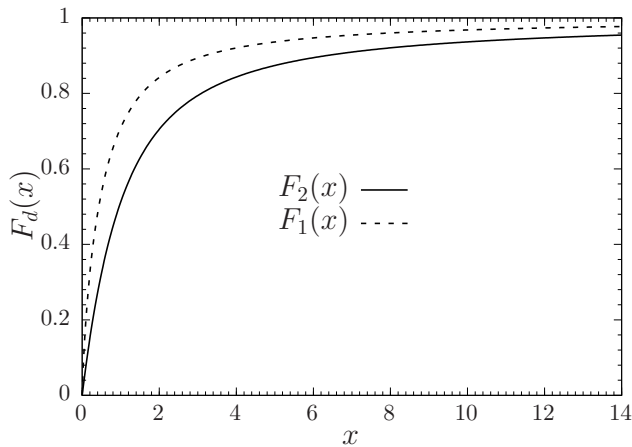}
\caption{\label{F}
Functions $F_2(x)$ and $F_1(x)$ of Eqs.~(\ref{F2D}) and (\ref{F1D}),
which determine the 2D and 1D exchange potential, respectively.
}
\end{figure}

Evaluation of  integrals in Eq.~(\ref{rofrp}) is straightforward, giving
\begin{equation}
\rho^\sigma(\rv_\|)  =  \frac{1}{2 \pi r_\|} 
\left\{  
\begin{array}{ll}
k_F^{\sigma}  J_1(k_F^{\sigma} r_\|), &d  =  2, \\[0.5em]
2 \sin (k_F^{\sigma} r_\|), &d  =  1,
\end{array}
\right.
\label{rofrpp}
\end{equation}
where $J_1(x)$ is the Bessel function of the first order.
With the use of Eqs.~(\ref{rofrpp}), the integral over $\rv'_\|$ in Eq.~(\ref{m1fin1}) can be
taken in special functions, resulting in
\begin{equation}
\begin{split}
  v^\sigma_x(r_\perp)   = 
  -  \int    \frac{F_d(k_F^{\sigma}|\rv_\perp  -  \rv'_\perp|) |\mu^\sigma_0(\rv'_\perp)|^2}{|\rv_\perp  - \rv'_\perp|}  d \rv'_\perp ,
\end{split}
\label{main15}
\end{equation} 
where
\begin{equation}
F_2(x)=1+\frac{ L_1(2 x)-I_1(2 x)}{x},
\label{F2D}
\end{equation}
\begin{equation}
F_1(x)=  \frac{1}{2\pi}  G_{2,4}^{2,2}\left[x^2  \left|
\begin{array}{l}
 \frac{1}{2},1 \\[0.25 em]
 \frac{1}{2},\frac{1}{2},-\frac{1}{2},0 \\
\end{array} \right.
\right],
\label{F1D}
\end{equation}
$L_1(x)$ and $I_1(x)$ are the first-order modified Struve and Bessel functions, respectively,
and
$G_{p,q}^{m,n}\left[x  \left|
\begin{array}{lll}
 a_1, &..., &a_p \\
 b_1, &..., &b_q \\
\end{array} \right.
\right]$ is the Meijer G-function \cite{Prudnikov,Mathematica}.
Functions $F_d(x)$ are plotted in Fig.~\ref{F}.

\section{Full 2(1)D confinement limit}
\label{STRICT}

We now take the limit of the strictly $d$-dimensional (zero-thickness) electron gas. 
Then $|\mu^\sigma_0(\rv_\perp)|^2=\delta(\rv_\perp)$,
and Eq.~(\ref{main15}) reduces to
\begin{equation}
\begin{split}
   v^\sigma_x(r_\perp)  =
-     \frac{F_d( k_F^{\sigma} r_\perp) }{r_\perp}.
\end{split}
\label{main17}
\end{equation} 
We emphasize that Eq.~(\ref{main17}) was obtained for the mathematical
idealization of an electron gas strictly confined to a plane (a straight line), as a limiting case
of Eq.~(\ref{main15}) for EG of a finite transverse extent.

The following zero-distance and asymptotic behaviour   
can be directly obtained from Eqs.~(\ref{main17}) and (\ref{F2D}) for the 2D case
\begin{align}
   &v^\sigma_x(z=0)  =  - \frac{8 k_F^{\sigma}}{3\pi}, \label{2D0}\\
   &v^\sigma_x(z\to \infty)  \to  - \frac{1}{z}+\frac{2}{\pi k_F^{\sigma} z^2} + ...  
\label{lim}
\end{align} 
and from Eqs.~(\ref{main17}) and (\ref{F1D}) for the 1D case
\begin{align}
   &v^\sigma_x(\rho\to 0) \to  \frac{k^\sigma_F}{\pi} [2 \log (k^\sigma_F \rho)+2 \gamma -3], \label{1D0}\\
   &v^\sigma_x(\rho\to \infty)  \to  - \frac{1}{\rho}+\frac{1}{\pi k_F^{\sigma} \rho^2} + ... ,
\label{lim1}
\end{align} 
where $\gamma\approx 0.5772$ is the Euler's constant. At large distances, 
at the both dimensionalities, the potential respects the 
asymptotic  $-1/r_\perp$.
\begin{figure} [h] 
\includegraphics[width=  \columnwidth, trim= 25 0 33 4, clip=true]{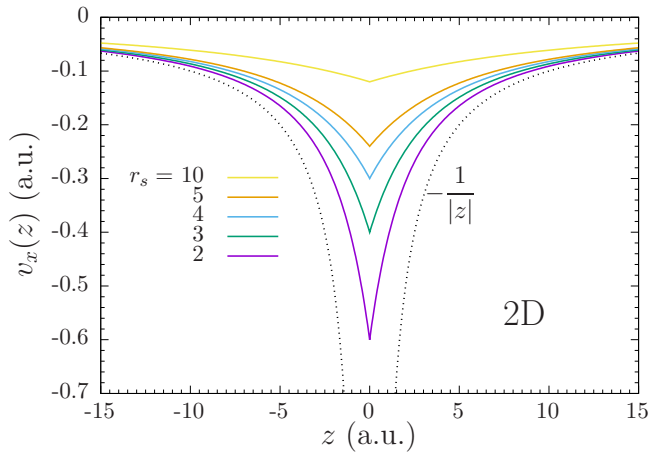}
\caption{\label{if}
EXX potential of the spin-unpolarized 2DEG, obtained with Eqs.~(\ref{main17}) and (\ref{F2D}), 
versus the distance from the EG plane $z$, with the density parameter $r_s=(\pi n)^{-1/2}$ 
changing from 10 a.u. (top) to 2 a.u. (bottom).
The dotted lines show the asymptotic $-1/|z|$. 
}
\end{figure}

\section{Discussion}
\label{DISC}

\begin{figure} [h] 
\includegraphics[width=  \columnwidth, trim= 26 0 35 4, clip=true]{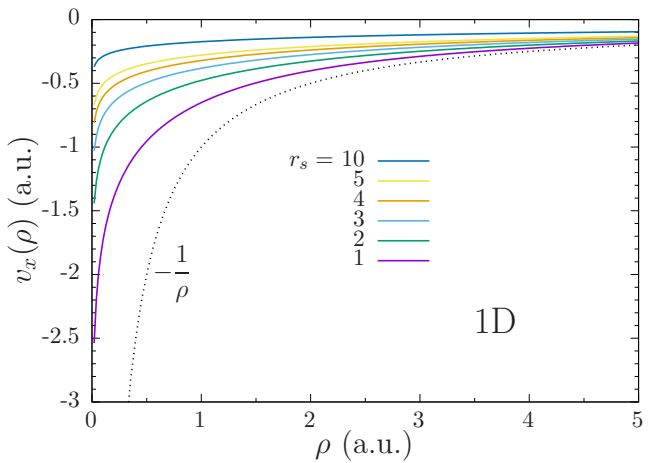}
\caption{\label{if1}
EXX potential of the spin-unpolarized 1DEG, obtained with Eqs.~(\ref{main17}) and (\ref{F1D}), versus the distance from the EG line $\rho$, with the density parameter $r_s=(2 n)^{-1}$ changing from 10 (top) to 1 a.u. (bottom).
The dotted line shows the asymptotic $-1/\rho$.
}
\end{figure}

It is known that, in the general case, EXX-OEP potential 
cannot be expressed in terms of the occupied states only, but all, the occupied and empty, states are involved,
and a complicated OEP integral equation must be solved to calculate this potential \cite{Sharp-53,Talman-76}.
It, therefore, may look surprising that a drastic simplification can be achieved in the case of 2(1)DEG with
one subband filled, leading to Eq.~(\ref{m1fin1}), the latter expressed in terms of the occupied states only and not involving
the OEP equation. To clarify this point, in Appendix \ref{FURTH2} we arrive at
Eq.~(\ref{m1fin1}) following the traditional path of working out the EXX-OEP potential in terms
of the eigenfunctions and the eigenenergies of all the states, seeing clearly how the simplifications arise
due to the specifics of this system. 

\begin{figure} [h] 
\includegraphics[width=\columnwidth, trim= 25 2 0 0, clip=true]{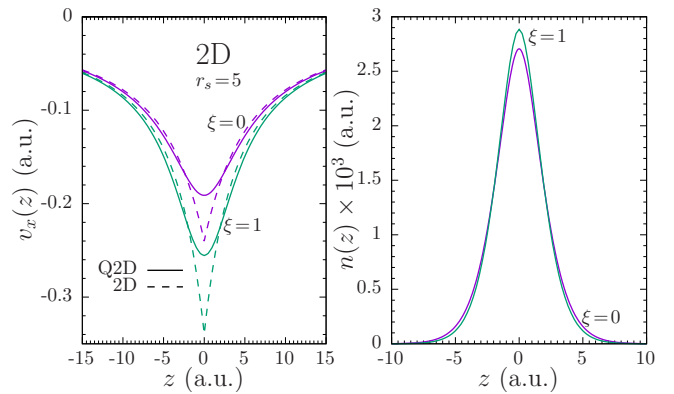}
\caption{\label{polvx} 
Left: Exchange potential as a function of the distance $z$ from the plane of the possitive background
of quasi-2D (solid lines) and 2D (dashed lines) EG,
for the spin-neutral ($\xi=0$) and fully spin-polarized ($\xi=1$) cases at the density-parameter $r_s=5$.
Right: The corresponding particle densities of the Q2D EG.
}
\end{figure}

\begin{figure} [h] 
\includegraphics[width=  \columnwidth, trim= 25 2 0 0, clip=true]{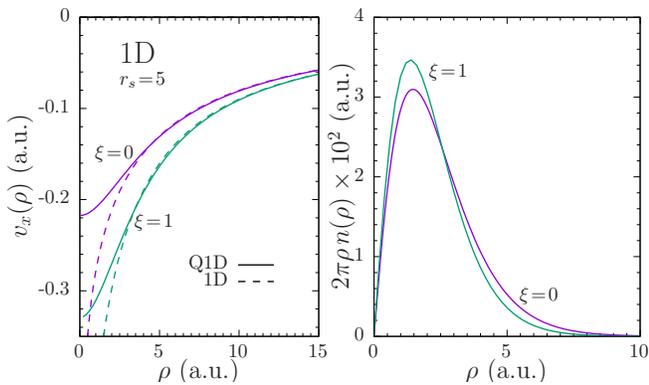}
\caption{\label{polvx1}
The same as Fig.~\ref{polvx}, but for 1DEG.
}
\end{figure}

In Figs.~\ref{if} and \ref{if1}, the EXX potentials of 2DEG and 1DEG, respectively, are 
plotted for a number of densities  as functions of the distance from the EG.
An important conclusion can be drawn from these figures together with 
the long-distance expansions (\ref{lim}) and (\ref{lim1}): The more dense is the EG, 
the sooner the  asymptotic is approached with the increase of the distance.
The characteristic distance scale is the inverse Fermi radius $k_F^{-1}$, which separates
two distinct  regions, the asymptotic one realizing at
$r_\perp \gtrsim k_F^{-1}$. 
We emphasize that in both regions there is no electron density, as is particularly
clear with the strictly 2D and 1D (zero-thickness) EG.  
\begin{figure} [h] 
\includegraphics[width=  \columnwidth, trim= 26 0 34 4, clip=true]{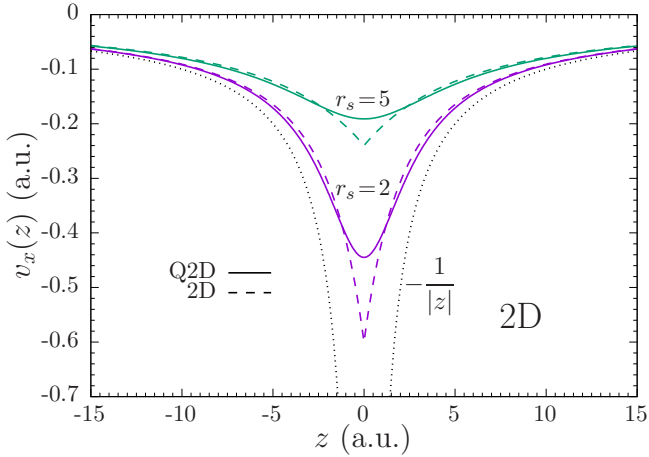}
\caption{\label{numvx}
The EXX potential of spin-neutral Q2D EG with one subband filled (solid lines)
compared to the analytical solution for the zero-thickness limit (dashed lines) for two
values of the density parameter $r_s=2$ and $5$. The dotted lines show the asymptotic $-1/|z|$.
}
\end{figure}  
\begin{figure} [h] 
\includegraphics[width=  \columnwidth, trim= 26 0 34 4, clip=true]{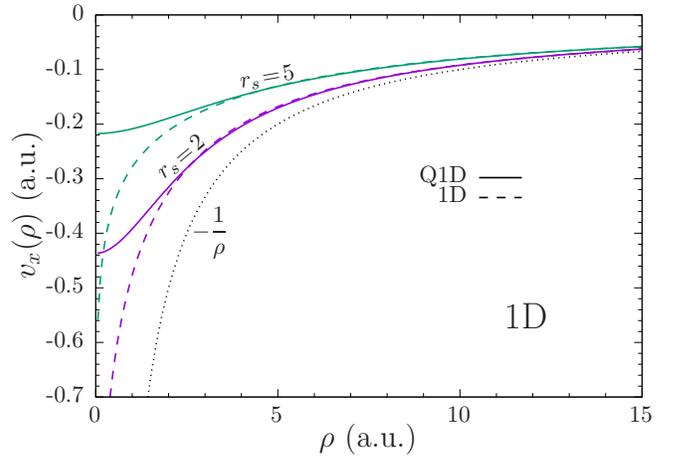}
\caption{\label{numvx1}
The same as Fig.~\ref{numvx} but for 1D case. 
}
\end{figure}
In Figs.~\ref{polvx} and \ref{polvx1} we present the exchange potential and the density for the       
spin-neutral and fully spin-polarized electron gas, in the 2D and 1D cases, respectively.

A natural question arises: How relevant are the solutions obtained for strictly
low-dimensional case to the realistic quasi-low-dimensional EG.
To answer this, we return to Eq.~(\ref{main15}) and solve the KS problem self-consistently
using  
\footnote{Although, separately the external and Hartree potentials  are infinite, 
their sum of Eq.~(\ref{vextH}) is finite.}
\begin{equation}
\begin{split}
&v_{ext,H}(\rv_\perp)=v_{ext}(\rv_\perp)+ v_H(\rv_\perp)= 2 \pi  \times \\ 
&  \left\{\begin{array}{ll}
  \displaystyle\int\limits_{-\infty}^\infty \left( |z| - |z-z'| \right) n(z')  d z' , & d=2, \\
 \! \! \displaystyle\int\limits_0^\infty \! \!  \rho' n(\rho') 
 \log \dfrac{2 \rho^2}{{\rho'}^2 \! + \! \rho^2 \! + \! |{\rho'}^2 \! - \! \rho^2|}   d\rho', & d=1.
\end{array}
\right.
\end{split}
\label{vextH}
\end{equation}
Results for the EXX potential presented in Figs.~\ref{numvx} and \ref{numvx1},
for the 2D and 1D cases, respectively, show that the zero-thickness limit of the EXX potential
is a very good approximation to that of the quasi-low-dimensional EG except for very short
distances from the system. At those distances, the electron-density is high 
(see Figs.~\ref{polvx} and \ref{polvx1}), 
the deviation in this region being not surprising, since the potential 
is not that in vacuum any more. 
Results of the calculations for the spin-polarized EG are presented in Figs.~\ref{polvx} and \ref{polvx1}.
\begin{figure} [h] 
\includegraphics[width=  \columnwidth, trim= 2 2 10 0, clip=true]{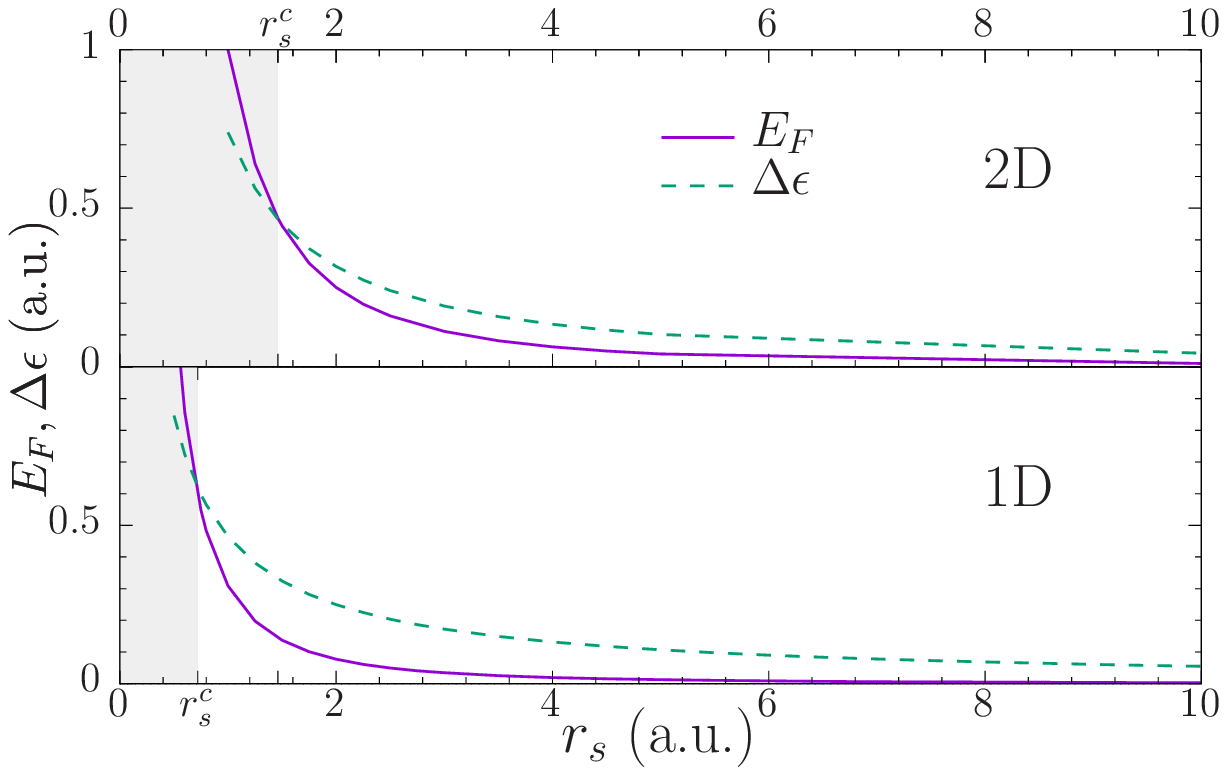}
\caption{\label{Efde}
Diagrams of  stability of Q2D (upper panel) and Q1D (lower panel) electron gases
with one subband occupied.
The Fermi energy $E_F=k_F^2/2$ and the distance $\Delta \epsilon$ between two lowest subbands  
are shown versus the density-parameter $r_s$.
At $r_s<r_s^c$ (shaded areas, $r_s^c \approx 1.46$ and $0.72$, in 2D and 1D cases, respectively),  
more than one subbands are filled, 
invalidating the results of this theory. The diagrams refer to the spin-unpolarized case.
}
\end{figure}
In Table \ref{tab} the low-lying eigenenergies of the
2D and Q2D EG are presented.
The deviation of the EXX potential from $-1/r_\perp$ in the non-asymptotic region causes
significant change in the spectra of the eigenenergies compared with the Rydberg series,
$-\frac{1}{2 n^2}$ and $-\frac{1}{2\left(n-\frac{1}{2}\right)^2}$, for 2D and 1D cases, respectively,
where $n=1,2,...$ \cite{Hassoun-81}.  
\begin{table} 
\begin{tabular}{lccccc}
\hline\hline
& Rydberg &  \multicolumn{2}{c}{$r_s=2$}   & \multicolumn{2}{c}{$r_s=5$} \\             
  $n$ & series   & 2D & Q2D & 2D & Q2D \\ \hline
   1  & 0.500     & 0.360    & 0.511  & 0.164  & 0.204 \\
   2  & 0.125     & 0.161    & 0.196  & 0.092  & 0.103 \\
   3  & 0.056     & 0.102    & 0.117  & 0.064  & 0.070 \\
   4  & 0.031     & 0.066    & 0.073  & 0.045  & 0.048 \\
   5  & 0.020     & 0.048    & 0.052  & 0.035  & 0.037 \\
   6  & 0.014     & 0.036    & 0.038  & 0.027  & 0.028 \\
\hline \hline
\end{tabular}
\caption{\label{tab} The first six eigenenergies (absolute values, a.~u.) of the EXX KS hamiltonian of 2D and Q2D EG for 
two  values of the density parameter $r_s=2$ and $5$,
compared with the Rydberg's series $1/(2 n^2)$.}
\end{table}

Having found  the analytical
EXX potential  of the strictly 2(1)DEG to be good approximations to the corresponding quasi-low-dimensional EG with
one filled subband, we need to establish when the latter regime actually holds.
This question is answered in the diagrams of the stability of the EG with  one  subband filled, 
presented in Fig.~\ref{Efde}. For $r_s>r_s^c$ ($r_s^c \approx 1.46$ and $0.72$ in 2D and 1D cases, respectively)
$E_F=k_F^2/2< \Delta \epsilon$, where $\Delta \epsilon$ is the subband gap, the state with one filled subband is stable.
Otherwise, at $r_s<r_s^c$, it is energetically preferable to start filling the second subband.
The latter regions, corresponding to high electron densities and to which our theory does not apply, are shaded at the diagrams. 

For the sake of completeness, we write down the total energy 
in the one occupied subband regime. Using the  DFT expression for  energy
\begin{equation}
\begin{split}
E &= \sum\limits_{i,\sigma} \epsilon^\sigma_i 
- \sum\limits_\sigma \int v^\sigma_{xc}(\rv)  
 n^\sigma(\rv) d\rv +  E_{xc}  \\
&-  \frac{1}{2 }  \int   v_H(\rv) n(\rv)  d\rv   +  \frac{1}{2}  \int  \frac{n^+(\rv) n^+(\rv')}{|\rv-\rv'|} d\rv d\rv',
\end{split}
\label{EDFT}
\end{equation} 
where the last term in Eq.~(\ref{EDFT}) is the energy of the background positive charge
$n^+(\rv)$,
by virtue  of Eqs.~(\ref{Ex2}) and (\ref{m1fin1}), we have for the energy per particle
\begin{equation}
\begin{split}
\epsilon &= \epsilon_K
+  \epsilon_{sub}
- \frac{1}{2 n}\sum\limits_\sigma \int v^\sigma_x(\rv_\perp)  
 n^\sigma(\rv_\perp) d\rv_\perp   \\
&  -  
\frac{1}{2 n}  \int  v_{ext,H}(\rv_\perp) \left[ n(\rv_\perp)+ n^+(\rv_\perp) \right] d\rv_\perp,
\end{split}
\end{equation} 
where the kinetic and the subband energies are
\begin{align}
\begin{split}
&\epsilon_K=\frac{1}{2 r_s^2}
\left\{
\begin{array}{ll}
1+\xi^2 , & d=2,\\
\pi^2 \left( 1+3 \xi^2\right)/48, & d=1,
\end{array}
\right.
\end{split}
\\
&\epsilon_{sub}=\frac{1}{2} \left[ \epsilon_0^\uparrow+\epsilon_0^\downarrow +
\xi \left( \epsilon_0^\uparrow-\epsilon_0^\downarrow \right) \right],
\end{align}
and $\xi=(n^\uparrow-n^\downarrow)/n$ is the spin polarization.
In the zero-thickness limit $v_{ext,H} \! = \! 0$, and  the exchange energy per particle becomes
$\epsilon_x=-\frac{2^{3/2}}{3\pi r_s} \left[ \left(1+\xi\right)^{3/2}+\left(1-\xi\right)^{3/2}\right]$
and $\epsilon_x=-\infty$, for 2D and 1D cases, 
in accord 
with Refs.~\onlinecite{Tanatar-89} and \onlinecite{Gold-90}, respectively. 

Our solutions, in particular, show that, while, in the general case,  EXX  requires
the knowledge of {\em all}, occupied and empty, states, and  solving of the optimized effective potential (OEP)
integral equation is necessary,
in the case of quasi-2(1)DEG with only one subband occupied, it is possible to avoid those complications,
still remaining within the exact theory. 
In this context, we note that the localized Hartree-Fock potential (LHF) \cite{Sala-01}, which has proven a useful concept requiring the occupied states only and involving no OEP equation in the general case,
yields results very close (often indistinguishable) to those of EXX and Hartree-Fock (HF) \cite{Sala-01,Nazarov-15-2}.
Conceptually, LHF potential can be constructed independently from HF and EXX \cite{Nazarov-13-2}, 
within the scheme
of the {\em optimized propagation} in time \cite{Nazarov-85},
and it is one of the realizations of the ``direct energy" potentials \cite{Levy-14,Levy-16}. 
This is, therefore, very instructive to learn
that, for  2(1)DEG with one filled subband, LHF coincides with both  EXX and Slater's potentials {\em exactly
up to a constant}, as we show in the next section. 

Here, it is instructive to draw the analogy with a singlet two-electron system
with only one state filled, for which all the three potentials coincide \cite{Nazarov-13-2,Nazarov-15-2}.
In the present case, although with an infinite number of electrons, the same property holds due to (i) the separation of the variables
in the two perpendicular directions and (ii) to the system being uniform (the potential being flat) in the parallel direction.

\section{Equivalence of EXX and LHF
for quasi-2(1)DEG with one filled subband}
\label{LHF}

The purpose of this section is to prove that, within the one-filled-subband regime
of 2(1)D electron gas, the localized Hartree-Fock (LHF) method and  and exact exchange (EXX) are exactly equivalent.

We start by reminding the basic facts on the LHF potential \cite{Sala-01}
within the framework of the optimized-propagation method (OPM) \cite{Nazarov-85,Nazarov-13-2,Nazarov-15-2}.
The LHF exchange potential $\tilde{v}^\sigma_x(\rv)$, experienced by electrons with the spin orientation $\sigma$,
is a solution to the equation 
\begin{widetext}
\begin{equation}
   \tilde{v}^\sigma_x(\rv) n^\sigma(\rv) =
\int  \left[   \tilde{v}^\sigma_x(\rv') -\frac{1}{|\rv-\rv'|}\right]  
|\rho^\sigma(\rv,\rv')|^2  d\rv' 
+ \int \frac{\rho^\sigma(\rv,\rv')\rho^\sigma(\rv',\rv'')\rho^\sigma(\rv'',\rv)}{|\rv'-\rv''|} d\rv' d\rv'',
\label{main1}
\end{equation}
\end{widetext}
with the spin-resolved density-matrix  defined as
\begin{equation}
\rho^\sigma(\rv,\rv')= {\sum\limits_{i}}^\sigma \phi_i(\rv) \phi_i^*(\rv'), 
\label{rofr}
\end{equation}
where the superscript  at the sum means that only the orbitals with the spin direction $\sigma$ are included.
For integegral number of particles, Eq.~(\ref{main1}) determines the potential up to the addition of an arbitrary constant $\tilde{v}^\sigma_x(\rv)\to \tilde{v}^\sigma_x(\rv)+c^\sigma$. \cite{Nazarov-15-2} A  relation
\begin{widetext}
\begin{equation}
\sum\limits_\sigma \left[\int \tilde{v}^\sigma_x(\rv) n^\sigma(\rv) d\rv + \frac{1}{2 } \int  \frac{|\rho^\sigma(\rv,\rv')|^2}{|\rv-\rv'|} d\rv d\rv' \right] 
 +   \frac{1}{2 } \int  v_H(\rv) n(\rv)  d\rv =0
\label{fconst}
\end{equation}
\end{widetext}
fixes this constant uniquely in the fully spin-polarized case, while in the presence
of electrons of both spin orientations, it makes only one of the constants
independent by fixing the quantity $c_\uparrow N_\uparrow+c_\downarrow N_\downarrow$,
where $N_\uparrow$ and $N_\downarrow$ are the number of electrons with spin up and spin down,
respectively. In both cases this fixes the total energy of the system,
the electronic part of which is equal in OPM to the sum of the single-particle eigenenergies \cite{Nazarov-15-2}.

For 2(1)DEG with only one subband filled
\begin{equation}
\rho^\sigma(\rv,\rv') = 
\mu^\sigma_0(\rv_\perp) {\mu^\sigma_0}^*(\rv_\perp') \rho^\sigma(\rv_\|-\rv'_\|),
\label{rrps}
\end{equation}
where $\rho^\sigma(\rv_\|)$ is given by Eq.~(\ref{rofrp}).
Substituting Eqs.~(\ref{rrps}) and (\ref{nnp}) into Eq.~(\ref{main1}) and 
{\em canceling out} $|\mu_{\sigma}(\rv_\perp)|^2$ {\em from both sides}, we can write
\begin{widetext}
\begin{equation}
\begin{split}
   n^\sigma \tilde{v}^\sigma_x(\rv_\perp) =
&- \int \frac{|\mu_{\sigma}(\rv_\perp')|^2 |\rho^\sigma(\rv_\|-\rv'_\|)|^2}{|\rv-\rv'|} d\rv' 
+\int  \tilde{v}^\sigma_x(\rv_\perp')   
|\mu_{\sigma}(\rv_\perp')|^2 |\rho^\sigma(\rv_\|-\rv'_\|)|^2  d\rv'
 \\
&+  \int  \frac{|\mu_{\sigma}(\rv_\perp')|^2 |\mu_{\sigma}(\rv_\perp'')|^2 
\rho^\sigma(\rv_\|  -  \rv_\|')\rho^\sigma(\rv'_\|  -  \rv_\|'')\rho^\sigma(\rv_\|''  -  \rv_\|)}{|\rv'-\rv''|} d\rv' d\rv'',
\end{split}
\label{m1}
\end{equation}
\end{widetext}
where we have explicitly written that the potential is a function of the perpendicular coordinate
only. We notice, and this is crucial to our derivation, that the second and the third 
terms in the right-hand side of Eq.~(\ref{m1}) are constants: The terms in question do not depend on $\rv_\perp$, and the apparent dependence
on  $\rv_{\|}$ is eliminated by the proper substitutions of the integration variables  $\rv_{\|}'$ and $\rv_{\|}''$.
Recalling that Eqs.~(\ref{main1}) are solvable up to arbitrary constants only, we can, therefore,  write
by virtue of Eq.~(\ref{m1})
\begin{equation}
   \tilde{v}^\sigma_x(\rv_\perp) = v^\sigma_x(\rv_\perp)+ c^\sigma,
\label{m1fin0}
\end{equation}
where $v^\sigma_x(\rv_\perp)$ is the EXX potential of Eq.~(\ref{m1fin1}).
Substituting Eq.~(\ref{m1fin0}) into Eq.~(\ref{fconst}), we have
\begin{widetext}
\begin{equation}
\sum\limits_\sigma c^\sigma N^\sigma =
-\sum\limits_\sigma \left[\int v^\sigma_x(\rv) n^\sigma(\rv) d\rv + \frac{1}{2 } \int  \frac{|\rho^\sigma(\rv,\rv')|^2}{|\rv-\rv'|} d\rv d\rv' \right] 
 -   \frac{1}{2 } \int  v_H(\rv) n(\rv)  d\rv.
 \label{cN}
\end{equation}
\end{widetext}

As already mentioned above, within the framework of OPM, the  electronic energy of a many-body 
system is a sum of the eigenvalues of its single-particle LHF hamiltonian
(i.e., it is a ``direct energy" potential \cite{Levy-14,Levy-16}). Therefore, we can write
\begin{equation}
E = \sum\limits_\sigma {\sum\limits_i}^\sigma \tilde{\epsilon}_i +  \frac{1}{2}  \int  \frac{n^+(\rv) n^+(\rv')}{|\rv-\rv'|} d\rv d\rv',
\label{ELHF}
\end{equation}
where the electrostatic energy of the positive background was added, as in the main text.
Since, due to Eq.~(\ref{m1fin0}), the LHF Hamiltonian differs from the EXX one by the constants $c^\sigma$ only,
we can write from Eq.~(\ref{ELHF})
\begin{equation}
E = \sum\limits_\sigma {\sum\limits_i}^\sigma \epsilon_i+ \sum\limits_\sigma c^\sigma N^\sigma
+  \frac{1}{2}  \int  \frac{n^+(\rv) n^+(\rv')}{|\rv-\rv'|} d\rv d\rv',
\label{ELHF2}
\end{equation}
where $\epsilon_i$ are the EXX eigenenergies.
Finally, substituting Eq.~(\ref{cN}) into  Eq.~(\ref{ELHF2}) and comparing with Eq.~(\ref{EDFT}),
we conclude that the LHF and EXX total energies exactly coincide.  

\section{Conclusions} 
\label{CONCL}

We have obtained explicit solutions to the problem
of the exact exchange -- optimized effective potential
of quasi-two- and one-dimensional electron gases with one subband filled in terms of the density.
It has been proven that the EXX potential, the localized Hartree-Fock potential, and the Slater's potential
all coincide with each other exactly (up to an arbitrary constant) for these systems.

By taking the limit of the zero-thickness [full 2(1)D confinement] of the respective electron gases,
we have found exact analytical solutions to the static EXX problem for 2(1)DEG.
We have identified a non-asymptotic regime, which realizes in the free space 
at distances less or comparable to the inverse Fermi radius of the electron gas.
While 
our solutions reproduce the before known asymptotic of the exchange potential at large distances,
at shorter distances the variance of the exchange potential  from its asymptotic strongly affects the low-lying
excited states, causing departure from the Rydberg series. 

Within a wide range of the densities of the electron gases, our analytical potentials
accurately approximate those of realistic quasi-2(1)D systems,
as demonstrated by the comparison to the results of the
self-consistent calculations  beyond the zero-thickness limit.
They, consequently, are expected to serve as efficient means to handle
the general problem of the exchange potential and image states in the  low-dimensional science.

As a by-product, our solutions extend a short list of analytical results known in the many-body physics.

\begin{acknowledgments}
Support from the Ministry of Science and Technology, Taiwan, Grant  No. \mbox{104-2112-M-001-007}, is acknowledged.                       
\end{acknowledgments}

\appendix

\section{Classical image potential of 2(1)D conductor}
\label{CLASS}

Let  2(1)D conductor occupy a plane (line) $\rv_\perp=\0v$ and let a test charge $Q$ be positioned
at $\rv_\perp=\Rv_\perp$ and $\rv_\|=\0v$. The field of the test charge will cause a change in the electron-density
distribution, which we denote $n_1(\rv_\perp)$. The latter is determined by the constancy of the potential
at the plane (line) of the conductor
\begin{equation}
\frac{Q}{\sqrt{R_\perp^2+\rv_\|^2}}- \int \frac{n_1(\rv_\|')}{|\rv_\|' -\rv_\| |} d\rv_\|' =0,
\end{equation}
by which we also fix the potential origin. Taking Fourier-transform with respect to 
$\rv_\|$ we have
\begin{equation}
n_1(\qv_\|)= Q \frac{v_{\qv_\|}(R_\perp)}{v_{\qv_\|}(0)},
\label{n1}
\end{equation}
where
\begin{equation}
v_{\qv_\|}(\rv_\perp)= \int \frac{e^{-\ii \qv_\|\cdot \rv_\|}}{\sqrt{\rv_\|^2+\rv_\perp^2}} d\rv_\|.
\label{vq}
\end{equation}
In real space Eq.~(\ref{n1}) yields for the density
\begin{equation}
n_1(\rv_\|)=  \frac{Q}{(2\pi)^d} \int  \frac{v_{\qv_\|}(R_\perp)}{v_{\qv_\|}(0)} e^{\ii \qv_\| \cdot \rv_\|}  d\qv_\|
\label{nE}
\end{equation}
and for its induced potential
\begin{equation}
\phi_{ind}(\rv_\|,\rv_\perp)= - \frac{Q}{(2\pi)^d} \int  \frac{v_{\qv_\|}(R_\perp) v_{\qv_\|}(r_\perp)}{v_{\qv_\|}(0)} e^{\ii \qv_\| \cdot \rv_\|}  d\qv_\|.
\label{fE}
\end{equation}
The total energy of the system, which, being the work required to move the test charge from $\Rv_\perp$ to
infinity and, therefore, is the image potential, is
\begin{equation}
E=Q \phi_{ind}(\rv_\|=\0v,\Rv_\perp)+ \frac{1}{2} \int \frac{n_1(\rv_\|) n_1(\rv'_\|)}{|\rv_\|-\rv'_\||} d\rv_\| d\rv'_\|, 
\end{equation}
which, by Eqs.~(\ref{nE}) and (\ref{fE}) can be rewritten as
\begin{equation}
\begin{split}
E & =- \frac{Q^2}{(2\pi)^d} \int  \frac{v^2_{\qv_\|}(R_\perp) }{v_{\qv_\|}(0)}  d\qv_\| + \frac{Q^2}{2 (2\pi)^d} \\ 
& \times \int  \frac{v^2_{\qv_\|}(R_\perp) }{v_{\qv_\|}(0)}  d\qv_\|  =
- \frac{Q^2}{2 (2\pi)^d} \int  \frac{v^2_{\qv_\|}(R_\perp) }{v_{\qv_\|}(0)}  d\qv_\|.
\end{split}
\label{EE}
\end{equation}

\subsection*{2D-case}
By Eq.~(\ref{vq}), $v_{\qv_\|}(R_\perp)=\frac{2\pi}{q_\|} e^{-q_\| R_\perp}$, and Eq.~(\ref{EE})
evaluates to
\begin{equation}
E= - \frac{Q^2}{4 R_\perp},
\end{equation}
which coincides with the result for a semi-infinite metal.

\subsection*{1D-case}
 By Eq.~(\ref{vq}), $v_{\qv_\|}(R_\perp)=2 K_0(q_\| R_\perp)$, 
where $K_n(x)$ is the modified Bessel function of the second kind. Since $K_0(0)=\infty$,
Eq.~(\ref{EE}) evaluates to
\begin{equation}
E= 0,
\end{equation}
in accordance that the notion of the strictly 1D electron gas is inherently inconsistent
and a finite width must be introduced for meaningful results \cite{Gold-90}.
These complications are not, however, relevant to the purposes of the present work.

\section{Further particulars of the derivation of Eq.~(\ref{m1fin1})}
\label{FURTH}

The derivation of Eq.~(\ref{m1fin1}) has been assuming that the system
remains that with one subband filled [and, therefore, Eq.~(\ref{Ex2}) holding] throughout the variational process.
Here we show that this, indeed, is the case. Let us write the variation of the exchange energy
\begin{equation}
\delta E_x = \int v_x(\rv_\perp) \delta n(\rv) d\rv,
\end{equation}
where, due to the symmetry of the problem, we have used the fact that $v_x$ is a function of the perpendicular coordinate only.
Therefore, the variontions of the density which avarage out to zero in the parallel to the EG direction (which is necessary for the conservation of the number of particles)  do not affect the the first order variation of $E_x$, the latter taken at the ground-state of the density. Therefore, with respect to finding the ground-state exchange potential for our systems, we need to consider the variations of the density which depend on $\rv_\perp$ only.

By the chain rule, this leads to the variation of the KS potential $v_s$ as a function of $\rv_\perp$ only, too.
Since the variables $\rv_\perp$ and $\rv_\|$ separate in KS equations and since the ground-state occupied spin-orbitals
(\ref{DMF}) factorize with the same perpendicular part for all of them, it is obvious, that, upon a variation of $v_s^\sigma(\rv)$,
the changed orbitals remain of the same form (\ref{DMF}) with one and the same, changed, perpendicular part. In other words,
through the variational procedure, the system remains that with only one subband filled.

\subsection{Proof of Eq.~(\ref{m1fin1}) in terms of the orbital wave-functions and eigenenergies}
\label{FURTH2}

Here we give an alternative proof of Eq.~(\ref{m1fin1}) which shows how the general EXX formalism 
leads to the Slater potential in the case of quasi 2(1)DEG with only one subband filled.
We can write
\begin{equation}
\frac{\delta E_x}{\delta v_s^\sigma(\rv)}= \int \frac{\delta n(\rv')}{\delta v_s^\sigma(\rv)} \frac{\delta E_x}{\delta n^\sigma(\rv')}  d\rv' =
\int \chi_s^\sigma(\rv,\rv') v_x^\sigma(\rv') d\rv',
\label{Ex333}
\end{equation}
where we have used the chain rule, the definitions of $v_x^\sigma(\rv)$ and of the KS spin-density-response function 
$\chi^\sigma_s(\rv,\rv')$, and the symmetry of the latter
in its two coordinate variables. The explicit expression of $\chi^\sigma_s(\rv,\rv')$ in terms of 
the KS wave-functions and eigenenergies is
\begin{equation}
\chi_s^\sigma(\rv,\rv')   =
\sum\limits_{\shortstack{\scriptsize{$i\in occ$} \\ \scriptsize{$j\not\in occ$}}}   
 \frac{\psi^{\sigma*}_i(\rv) \psi_j^\sigma(\rv)  \psi^{\sigma*}_j(\rv') \psi^\sigma_i(\rv')} { \epsilon_i - \epsilon_j  } + c.c.   
\label{chi_KS_s}
\end{equation}
In the case of 2(1)D EG with the flat in-plane potential, the variables $\rv_\|$ and 
$\rv_\perp$ separate in the KS equations, $i$ ($j$) becomes a composite index
$i=(\kv_{\| i},n_i)$, where $\kv_\|$ is the parallel momentum, $n$ indexes the transverse bands,
and 
\begin{equation}
\epsilon^\sigma_i= \frac{k_{\|i}^2}{2}+ \lambda_{n_i}^\sigma,
\label{ee}
\end{equation} 
\begin{equation}
\phi_i^\sigma(\rv)=\frac{e^{\ii \kv_{\| i} \cdot \rv_\|}}{\Omega^{1/2}} \mu^\sigma_{n_i}(\rv_\perp),
\label{DMF2}
\end{equation}  
where $\mu^\sigma_{n_i}(\rv_\perp)$ are the wave-functions of the $(3-d)$-dimensional motion in 
the potential $v_s^\sigma(\rv_\perp)$, and $\lambda_{n_i}$ are the corresponding eigenenergies.
We substitute Eqs.~(\ref{chi_KS_s})-(\ref{DMF2}) into Eq.~(\ref{Ex333}) and
note that, by the symmetry of the problem, the left-hand side of Eq.~(\ref{Ex333})  depends on $\rv_\perp$ only and,
therefore, we can  average Eq.~(\ref{Ex333})  in $\rv_\|$ over $\Omega$ without changing it.
Doing this and integrating over $\rv'_\|$ on the right-hand side with account of $v_x^\sigma(\rv')$ being a function of $\rv'_\perp$ only, we see that only $\kv_{\| j}=\kv_{\| i}$ contribute, leading to
\begin{equation}
\frac{\delta E_x}{\delta v_s^\sigma(\rv)}= n^\sigma
\int \chi_s^\sigma(\rv_\perp,\rv'_\perp) v_x^\sigma(\rv'_\perp) d\rv'_\perp,
\label{Ex0}
\end{equation}
where
\begin{equation}
\chi_s(\rv_\perp,\rv'_\perp)=  
\sum\limits_{n=1}^\infty 
\frac{
\mu^{\sigma*}_0(\rv'_\perp) 
 \mu^\sigma_n(\rv'_\perp) 
\mu^\sigma_0(\rv_\perp) 
\mu^{\sigma*}_n(\rv_\perp)}
{ \lambda_0^\sigma - \lambda_n^\sigma} + c.c. ,
\label{ch1}
\end{equation}
and we have taken account of the fact that only one state with 
the wave-function $\mu_0^\sigma(\rv_\perp)$ is occupied in the transverse direction.

On the other hand, we can evaluate $\delta E_x/\delta v_s^\sigma(\rv)$ directly.
By Eqs.~(\ref{Exg}) and (\ref{ro})
\begin{equation}
E_x= -\frac{1}{2} \sum\limits_\sigma \sum\limits_{\shortstack{\scriptsize{$(i,\! \sigma) \! \in \! occ$} \\ \scriptsize{$(j,\! \sigma) \! \in \! occ$}}} \int \frac{\phi^\sigma_i(\rv) \phi_j^{\sigma*}(\rv) \phi_j^\sigma(\rv') \phi_i^{\sigma*}(\rv')}{|\rv-\rv'|} d\rv d\rv'.
\label{Exffff}
\end{equation}
Then, using the chain rule,
\begin{equation}
\begin{split}
\frac{\delta E_x}{\delta v_s^\sigma(\rv)} \! = \! \! \! \! \!
\sum\limits_{m\in occ} \int \! \left[ \frac{\delta E_x}{\delta \phi_m^\sigma(\rv')} \frac{\delta \phi_m^\sigma(\rv')}{\delta v_s^\sigma(\rv)} \! + \!
\frac{\delta E_x}{\delta \phi_m^{\sigma*}(\rv')} \frac{\delta \phi_m^{\sigma*}(\rv')}{\delta v_s^\sigma(\rv)}\right] \! d\rv'. 
\end{split}
\label{Exfi}
\end{equation}

Due to Eq.~(\ref{Exffff})
\begin{equation}
\frac{\delta E_x}{\delta \phi_m^{\sigma}(\rv')} \! = \!
- \! \! \! \! \! \sum\limits_{(i,\sigma)\in occ} \!
\int \! \frac{\phi^\sigma_i(\rv'') \phi_m^{\sigma*}(\rv'') \phi_i^{\sigma*}(\rv') }{|\rv'-\rv''|} d\rv'',  (m,\sigma) \! \in \! occ.
\label{s1}
\end{equation}
By the use of the perturbation theory, we can write
\begin{equation}
\frac{\delta \phi_m(\rv')}{\delta v_s^\sigma(\rv)}= 
\sum\limits_{l\ne m} \frac{\phi_l^\sigma(\rv') \phi_m^\sigma(\rv) \phi_l^{\sigma*}(\rv) }{\epsilon_m^\sigma-\epsilon_l^\sigma}.
\label{s2}
\end{equation}
Substitution of Eqs.~(\ref{s1}) and (\ref{s2}) into Eq.~(\ref{Exfi}) gives
\begin{widetext}
\begin{equation}
\frac{\delta E_x}{\delta v_s^\sigma(\rv)}=- \sum\limits_{\shortstack{\scriptsize{$(m,\! \sigma) \! \in \! occ$} \\ \scriptsize{$(i, \sigma)  \in \! occ$}}} \
 \sum\limits_{l\ne m} \int \left[   
\frac{\phi^\sigma_i(\rv'') \phi_m^{\sigma*}(\rv'') \phi_i^{\sigma*}(\rv') \phi_l^\sigma(\rv') \phi_m^\sigma(\rv) \phi_l^{\sigma*}(\rv)}{|\rv'-\rv''| (\epsilon_m^\sigma-\epsilon_l^\sigma)}  
+
c.c. \right]  d\rv' d\rv''.
\label{Ex22}
\end{equation}

Further, substituting Eqs.~(\ref{ee}) and (\ref{DMF2}) into Eq.~(\ref{Ex22}), we have
\begin{equation}
\begin{split}
&\frac{\delta E_x}{\delta v_s^\sigma(\rv)}=- \frac{1}{\Omega^3}  
\sum\limits_{\shortstack{\scriptsize{$|\kv_{\| i}| \le k^\sigma_F$} \\ \scriptsize{$|\kv_{\| m}| \le k^\sigma_F$}}} \ \ 
\sum\limits_{(\kv_{\| l},n_l)\ne (\kv_{\| m}, 0)} \\
& \int \left[   
\frac{
e^{\ii (\kv_{\| i}-\kv_{\| m} )\cdot \rv''_\|} \mu^\sigma_0(\rv''_\perp) 
\mu^{\sigma*}_0(\rv''_\perp)
\mu^{\sigma*}_0(\rv'_\perp) 
e^{\ii (\kv_{\| l}-\kv_{\| i} )\cdot \rv'_\|} \mu^\sigma_{n_l}(\rv'_\perp) 
e^{\ii (\kv_{\| m}-\kv_{\| l}) \cdot \rv_\|} \mu^\sigma_0(\rv_\perp) 
\mu^{\sigma*}_{n_l}(\rv_\perp)}
{ 
\left( \frac{k_{\|m}^2}{2}+ \lambda_0^\sigma - \frac{k_{\|l}^2}{2}- \lambda_{n_l}^\sigma \right) |\rv'-\rv''|} 
+
c.c. \right]  d\rv' d\rv'',
\end{split}
\label{Ex222}
\end{equation}
where, again, it has been taken into account that the occupied orbitals have only one transverse band $\mu^\sigma_0(\rv_\perp)$.

Similar to the above, we average Eq.~(\ref{Ex222}) over $\rv_\|$ without changing it.
This leads to only the terms with $\kv_{\| l}=\kv_{\| m}$ remaining, and  gives
\begin{equation}
\frac{\delta E_x}{\delta v_s^\sigma(\rv)}=- \frac{1}{\Omega^3}  
\sum\limits_{\shortstack{\scriptsize{$|\kv_{\| i}| \le k^\sigma_F$} \\ \scriptsize{$|\kv_{\| m}| \le k^\sigma_F$}} } \ 
\sum\limits_{n=1}^\infty 
 \int \left[   
\frac{ |\mu^\sigma_0(\rv''_\perp)|^2
\mu^{\sigma*}_0(\rv'_\perp) 
 \mu^\sigma_n(\rv'_\perp) 
\mu^\sigma_0(\rv_\perp) 
\mu^{\sigma*}_n(\rv_\perp) e^{\ii (\kv_{\| m}-\kv_{\| i} )\cdot (\rv'_\|-\rv''_\|)}}
{ 
\left( \lambda_0^\sigma - \lambda_n^\sigma \right) |\rv'-\rv''|} 
+
c.c. \right]  d\rv' d\rv'',
\label{Ex2222}
\end{equation}
or, according to the definition (\ref{rofrp}),
\begin{equation}
\frac{\delta E_x}{\delta v_s^\sigma(\rv)}=- \frac{1}{\Omega}  
\sum\limits_{n=1}^\infty 
 \int  \left[ 
\frac{ |\mu^\sigma_0(\rv''_\perp)|^2 
\mu^{\sigma*}_0(\rv'_\perp) 
 \mu^\sigma_n(\rv'_\perp) 
\mu^\sigma_0(\rv_\perp) 
\mu^{\sigma*}_n(\rv_\perp) |\rho^\sigma(\rv'_\|-\rv''_\|)|^2}
{ 
\left( \lambda_0^\sigma - \lambda_n^\sigma \right) |\rv'-\rv''|}  +c.c. \right] d\rv' d\rv'',
\label{Ex22222}
\end{equation}
or, by Eq.~(\ref{ch1}) and integrating over $\rv'_\|$,
\begin{equation}
\frac{\delta E_x}{\delta v_s^\sigma(\rv)}= \! - \!  
 \int  \! d\rv'_\perp \chi_s^\sigma(\rv_\perp,\rv'_\perp)
\! \! \int \frac{ |\mu^\sigma_0(\rv''_\perp)|^2  
|\rho^\sigma(\rv'_\| \! - \! \rv''_\|)|^2}
{ 
|\rv'-\rv''|}   d\rv''.
\label{Ex222222}
\end{equation}
Combining Eqs.~(\ref{Ex0}) and (\ref{Ex222222}), we have
\begin{equation}
\int d\rv'_\perp \chi_s^\sigma(\rv_\perp,\rv'_\perp) 
 \left[ n^\sigma v_x^\sigma(\rv'_\perp)  + 
 \int \frac{ |\mu^\sigma_0(\rv''_\perp)|^2  
|\rho^\sigma(\rv'_\|-\rv''_\|)|^2}
{ |\rv'-\rv''|}    d\rv'' \right]=0 .
\label{fin}
\end{equation}

For Eq.~(\ref{fin}) to hold, the expression in the square brackets must be a constant,
which retrieves the exchange potential of Eq.~(\ref{m1fin1}) up to an arbitrary constant.
\end{widetext}


%

\end{document}